\documentclass[aps,pra,a4paper,showpacs,onecolumn]{revtex4}

\usepackage[latin1]{inputenc}
\usepackage{amsmath}
\usepackage{amsfonts}
\usepackage{amssymb}
\usepackage{bm}
\usepackage{graphicx}
\usepackage{epsfig}
\usepackage{subfigure}
\usepackage{setspace}

\def\be{\begin{equation}}
\def\ee{\end{equation}}
\def\bea{\begin{eqnarray}}
\def\eea{\end{eqnarray}}

\def\bfE{\mathbf{E}}
\def\bfP{\mathbf{P}}
\def\bfr{\mathbf{r}}

\def\calN{\mathcal{N}}
\def\c{C}

\def\Hga{$6^1S_0$}
\def\Hgb{$6^3P_1$} 
\def\Hgc{$7^1S_0$} 
\def\i{i}
\def\ii{ii}

\def\eff{\hbox{\scriptsize{eff}}}
\def\Oeff{ \Omega_{\hbox{\scriptsize{eff}}} }
\def\half{\frac{_1}{^2}}

\begin{document}
\setstretch{2}

\title{Preparation of nondegenerate coherent superpositions \\ 
in a three-state ladder system assisted by Stark Shifts}
\date{\today}
\author{N. Sangouard}
\email{nicolas.sangouard@physik.uni-kl.de}
\affiliation{Fachbereich Physik, Universit\"at Kaiserslautern,
67653, Kaiserslautern, Germany}
\author{L.P. Yatsenko}
\altaffiliation{Permanent address: Institute~of~Physics, Ukrainian Academy
of Sciences, ~prospect~Nauki~46,~Kiev-22,~252650,~Ukraine}
\author{B.W. Shore}
\altaffiliation{Permanent address: 618 Escondido Cir.,~Livermore~CA}
\author{T. Halfmann}
 \homepage{http://www.quantumcontrol.de}
\affiliation{Fachbereich Physik, Universit\"at Kaiserslautern,
67653, Kaiserslautern, Germany}

\begin{abstract}
We propose a technique to prepare coherent superpositions of two 
nondegenerate quantum states
in a three-state ladder system, driven by two simultaneous fields
near resonance with an intermediate state. The technique, of
potential application to enhancement of nonlinear processes, uses
adiabatic passage assisted by dynamic Stark shifts induced by a
third laser field. The method offers significant advantages over
alternative techniques: 
(\i) it does not require laser pulses of specific shape and duration and
(\ii)  it requires less intense fields
than schemes based on two-photon excitation with non-resonant
intermediate states. We discuss possible experimental
implementation for enhancement of frequency conversion in mercury atoms.
\end{abstract}

\maketitle


\section{Introduction}

Efficient mechanisms to generate coherent superpositions of
quantum states are central to a rich variety of applications in
modern quantum physics. Quantum logic gates, i.e. the key
components of a quantum computer%
~\cite{Galindo_rmp02},  
rely on superpositions of two degenerate quantum states as qubits%
~\cite{Kis_pra02,Sangouard_arxiv05}.  
Quantum tunnelling and localization in a
double well potential can be controlled%
~\cite{Sangouard_prl04} 
by techniques that require preparation of coherent superpositions.
Numerous authors have noted that nonlinear optical processes, e.g. resonantly enhanced
frequency mixing in atomic vapours, can be significantly improved%
~\cite{Jain_prl96,Luk_prl98,Andrew_ol99,Mys_oc02,Kor_epjd03,Ric_oc03} 
by preparing the nonlinear optical medium in a coherent
superposition of nondegenerate quantum states. 
Whereas numerous techniques exist to prepare
coherent superpositions of degenerate states, as needed for qubits in quantum
computing, nonlinear optics still hold challenges. For example,
frequency conversion to short wavelength radiation involves
high-lying states, requiring excitation by multi-photon
transitions. 

Procedures based on adiabatic passage driven by
coherent interactions%
~\cite{Vit_arpc01} 
provide reliable and
robust tools for the creation of superpositions. Unlike diabatic
techniques, which rely on precise control of individual pulses,
adiabatic processes are insensitive to small variations of pulse
duration and peak intensity. High-intensity laser systems, commonly
used for efficient frequency conversion,
exhibit fluctuations in intensity and other parameters. Thus
robust techniques are most appropriate to support nonlinear
optical processes  driven by the high-intensity lasers. 

\begin{figure}[!ht]
\includegraphics[scale=0.5]{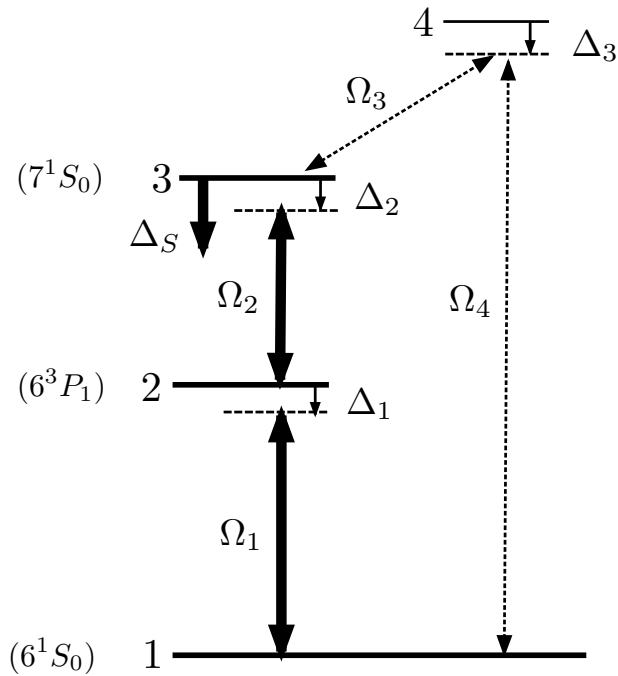}
\caption{ 
 Three-state system in ladder linkage,
interacting with two pulsed laser fields, indicating
 the detunings $\Delta_j$,
 the Rabi frequencies $\Omega_j$ and the Stark shift $\Delta_S$. 
 Dotted lines show transitions for four-wave mixing, through a nonresonant state 4.
 To the left are the spectroscopic labels appropriate to implementation with mercury.
 }
\label{scheme_config}
\end{figure}

In what follows we will discuss adiabatic passage in a
three-state system involving two near-resonant laser pulses. We
assume that initially the atom is in the ground state 1
 of energy $E_1$ and that the first laser field, of frequency $\omega_1$, 
links this with an excited state 2 of energy $E_2,$ while the second field, 
of frequency $\omega_2$, links this state with a final target state 3 of energy $E_3.$ 
We assume that the two fields, though near resonant with the specified transitions, are far from resonance with any other transition. The relative ordering of the third-state energy $E_3$ is not significant; we shall assume that it lies above $E_2,$ as is appropriate for application to nonlinear optics, so that the linkages form a ladder.
Figure  \ref{scheme_config} 
shows this linkage pattern, along with spectroscopic labels 
appropriate to implementation in mercury.

Our modeling is based upon the three-state time-dependent Schr\"odinger equation
in the rotating wave approximation (RWA)%
~ \cite{Shore_ny90}, 
for which the Hamiltonian is
\begin{equation}
\label{Hamiltonian-3state}
H(t) =\hbar
\left[ \begin{array}{ccc}
      0 & \half \Omega_{1}(t)  & 0 \\
     \half  \Omega_{1}(t)  & \Delta_2 & \half \Omega_{2}(t)  \\
      0 & \half \Omega_{2}(t)  & \Delta_2+\Delta_3-\Delta_S(t)
\end{array} \right].
\end{equation}
The elements of this RWA Hamiltonian are the detunings
\be
\Delta_2 = (E_2 - E_1)/\hbar - \omega_1,
\qquad
\Delta_3 = (E_3 - E_2)/\hbar - \omega_2.
\ee
and the pulsed Rabi frequencies
$\Omega_1(t)=d_{12}\mathcal{E}_1(t)/\hbar$ 
and
$\Omega_2(t)=d_{23}\mathcal{E}_2(t)/\hbar$
that parametrize the laser-induced excitation interaction 
 with the dipole-transition moments $d_{nm}$ of the transition
between states $n$ and $m$ and the electric field envelopes 
$\mathcal{E}_1(t)$ and ${\mathcal{E}}_2(t)$.
A third laser, nonresonant with either of these transitions, induces dynamic Stark shifts of the energies. For the ladder-like linkage the shift is expected to be largest for state 3; we denote that shift as $\Delta_S$ and we neglect shifts of the other states.

In the RWA the statevector has the expansion
\be
\Psi(t) = C_1(t) \psi_1 + C_2(t) e^{-i\omega_1 t} \psi_2 
+ C_3(t) e^{-i(\omega_1+\omega_2) t} \psi_3 
\ee
where $\psi_n$ is a unit vector representing quantum state $n$.
Our principal objective is to create, for times $t$ later than the conclusion
of a pulse sequence at $t_f$, an equal-probability superposition
of states $\psi_1$ and $\psi_3$, specifically
\begin{equation}
\label{final_super}
\Psi(t)=\frac{1}{\sqrt{2}}
\left[\psi_1+ e^{-i\varphi} e^{-i(\omega_1 + \omega_2) t } \psi_3\right] ,
 \qquad t > t_f
\end{equation}
where $\varphi$ is a time independent phase defining the relative
sign of the superposition.  This particular
superposition, with equal probabilities 
$P_n(t)=|\c_n(t)|^2=|\langle \psi_n| \Psi(t) \rangle |^2$ 
of the two constituent nondegenerate states,
provides the basis of a technique referred to as ``nonlinear
optics at maximum coherence''%
~\cite{Kor_epjd03}, 
that substantially improves
the efficiency of nonlinear frequency
conversion processes.  
The following section reviews
 how the efficiency of four-wave
mixing is enhanced by preparing the atomic medium in the coherent
superposition of eqn. (\ref{final_super}) with equal probability
amplitudes. 
We review in section \ref{sec-background} two
techniques recently suggested to generate coherent superpositions
of this form: 
(\i) fractionally-completed stimulated Raman
adiabatic passage (F-STIRAP)  and 
(\ii) half-completed Stark-chirped
rapid adiabatic passage (half-SCRAP). 
Section \ref{sec-SACS}
describes our proposed method, Stark-assisted coherent superposition (SACS),
in which two near-resonant fields,
accompanied by a laser-induced AC Stark shift, produce adiabatic
passage into the desired superposition. 
Section \ref{sec-simulation} 
illustrates different techniques by presenting simulations of  excitation of mercury vapor, a medium of significant interest
for applications in frequency conversion.

\section{Enhancement of four-wave mixing by maximum coherence}\label{four_wave_mixing}

The starting point for nonlinear optical phenomena is the wave equation describing the propagation of the electric field $\bfE$ through matter. The effects of matter are incorporated into a polarization $\bfP$ that serves as an inhomogeneous term in the wave equation,
\bea
& &\left[ \nabla ^2 - {1 \over c^2} {\partial ^2 \over \partial t^2}\right]
{\bfE}(t,\bfr) = \mu_0 {\partial ^2 \over \partial t^2} {\bfP}(t,\bfr).
\label{12.2-25}
\eea
We idealize the medium as a uniform distribution of identical motionless atoms, of number density
$\calN$. Then the polarization is the $\calN$ time the expectation value $\langle \mathbf{d}(t,\bfr) \rangle$ of the single-atom dipole moment:
 \be
{\bfP}(t,\bfr) = \calN \langle {\mathbf{d}}(t,\bfr) \rangle. 
\ee
In the following we idealize the laser fields  as plane
waves traveling along the $z$ axis. We suppose that there are
several such fields, each characterized by a common polarization
vector $\mathbf{e}$ but different carrier frequency $\omega_k$. We
express the resulting electric vector as
 \be {\bfE}(t,\bfr) = \sum_k \mathbf{e} \,
{\cal{E}}_k(t') \cos(\omega_k t' + \phi_k) 
\ee 
where $t' = t - z/c.$ The presence of the laser fields causes changes of the
material. With these changes come changes in the individual dipole
expectation values and hence of the polarization $\bfP.$ We will 
discuss a four-wave mixing process in an
atomic gas, as are typical for the generation of
short-wavelength radiation, i.e. vacuum- or extreme-ultraviolet. 
For application to four-wave mixing in mercury, 
the three states depicted in Fig. \ref{scheme_config}
 are: (1) \Hga , (2) \Hgb and (3)  \Hgc.

For the four-wave mixing processes, two fields at frequencies
$\omega_1$ and $\omega_2$ create an induced dipole moment. This 
dipole moment, together with a third field of frequency
$\omega_3$ far from resonance with any state $\psi_4,$
combine to produce a dipole moment that varies at the frequency
$\omega_4=\omega_1+\omega_2+\omega_3$. Following the
definition of the expectation value of the dipole moment
$
\langle \mathbf{d}(t') \rangle= \langle \Psi(t')|\mathbf{d}|
\Psi(t') \rangle, 
$
it can be expressed as a function of
the probability amplitude
 \bea \langle {\mathbf{d}}(t') \rangle
\cdot \mathbf{e}&=& 2 \hbox{Re}\left[
 \c^{\ast}_1(t') \c_2(t') d_{12} e^{i\omega_1 t'}
+ \c^{\ast}_2(t') \c_3(t') d_{23} e^{i\omega_2 t'} 
 \right. \nonumber \\ && \quad \left.
+ \c^{\ast}_3(t') \c_4(t') d_{34} e^{i\omega_3 t'}
+ \c^{\ast}_4(t')\c_1(t') d_{41} e^{i\omega_4 t'}
\right]. 
\eea 
The last term serves
as the source of an electric field at the frequency $\omega_4.$
For a detuning $\Delta_4=(E_4-E_3)/\hbar-\omega_3$ much larger
than the Rabi frequency $\Omega_3,$ one can adiabatically
eliminate the state $\psi_4$%
~\cite{Shore_ny90}.  
The amplitude
$\c_4(t')$ is then given by 
\be
\c_4(t')=\frac{\Omega_3(t')}{2\Delta_4}
\c_3(t')+\frac{\Omega_4(t')}{2\Delta_4} \c_1(t'). 
\ee 
When
$\Omega_4$ is small compared to $\Omega_3$, as it is initially, the
source term of the electric field at frequency $\omega_4$ is 
\be
 \label{source_term4}
 \mathcal{P}_4(t') = 
 \c^{\ast}_3(t') \c_1(t') \mathcal{E}_3(t') \frac{d_{23} d_{41} }{2\hbar \Delta_4}
e^{i\omega_4 t'} 
\ee 
where $\mathcal{P}_j(t)$,  for $j$=(1,2,3,4),  is defined from the expression
 \be
  \mathbf{P}(t')=2\mathbf{e}\hbox{Re}\left[\sum_{j=1}^{4}
\mathcal{P}_j(t') e^{i\omega_j t'}\right]. 
\ee 
These successive equations
show  that the induced polarization, and hence the efficiency of any
subsequent nonlinear optical process,  depends on the product
$\c^{\ast}_3(t') \c_1(t')$, i.e. on the coherence, established between
the ground and the excited state. Laser fields typically exhibit
inhomogeneities in time, space or phase. If there is a
dynamical contribution to the phase, these inhomogeneities will,
when averaged, tend to reduce any coherence to nearly zero.
To generate the desired radiation at the frequency $\omega_4$ it
is thus crucial that the relative phase of the components 
$\c_1(t)$ and $\c_3(t)$
 in the superposition state of eqn. (\ref{final_super}) does not
depend on the time integrated Rabi frequencies, as dynamical phases
do. If this condition is fulfilled, the source term  of eqn.
(\ref{source_term4}) reaches a maximum value when the atomic
coherence is maximum, i.e. when 
$|\c_1(t)|=|\c_3(t)|$. The polarization,
the efficiency of four-wave mixing and thus the intensity of the
generated radiation at frequency $\omega_4$ will be enhanced, if
the atomic medium is prepared in maximum coherence (\ref{final_super}).

\section{Adiabatic preparation of coherent superpositions}  \label{sec-background}

\begin{figure}[!ht]
\includegraphics[scale=0.35]{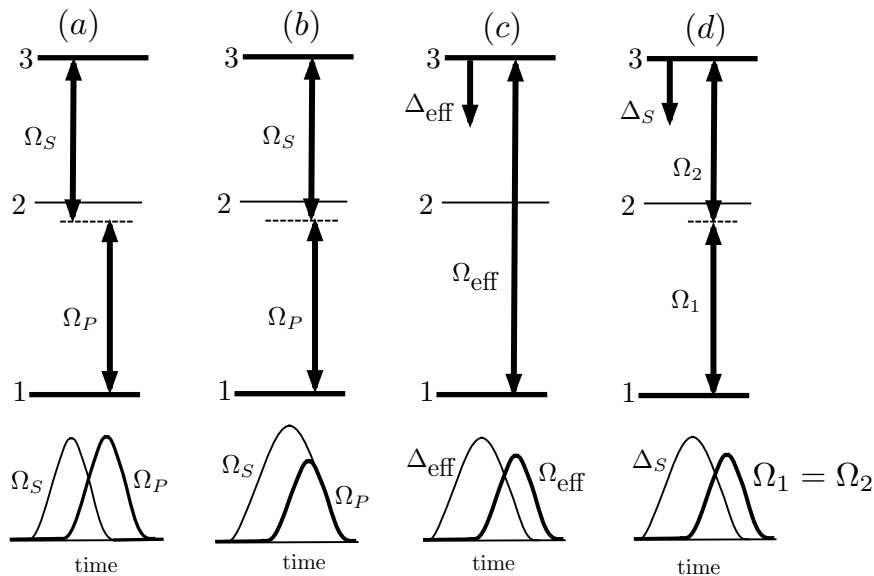}
\caption{Three-level system in
ladder-type configuration for the specific
adiabatic excitation processes: 
(a) STIRAP,
(b) F-STIRAP
(c) half-SCRAP, 
(d)  SACS.
Below the linkage patterns are indicators of the time
dependence of the pulses.
 } \label{scheme_comparison}
\end{figure}

As will be discussed below, to produce this superposition we make use of adiabatic states $\Phi_k(t)$
 and adiabatic eigenvalues, $\lambda_k(t)$, defined as instantaneous solutions
 to the eigenvalue equation of the RWA Hamiltonian
\be
 H(t) \Phi_k(t) = \hbar \lambda_k(t) \Phi_k(t).
\ee
Prior to describing the proposed technique, we comment on two common
adiabatic techniques, first intended for complete population transfer but later applied
to the preparation of superposition states.
Figure \ref{scheme_comparison} illustrates the three schemes.

\subsection{STIRAP and F-STIRAP}

Stimulated Raman adiabatic passage
(STIRAP)%
~\cite{Gaubatz_jcp90,Bergmann_rmp98}  
is well established as an
appropriate tool to drive complete population transfer in a
three-state lambda-type system%
~\cite{Vit_arpc01}.  
With suitable modification%
~\cite{Marte_pra91,Weitz_pra94,Goldner_prl94,
Lawal_prl94,Vitanov_jpb99,Sautenkov_pra04}, 
the technique provides a means of creating superpositions of
initial and final states. Essentially the evolution is
adiabatically stopped just when half of the population is transferred. We refer to this as fractional STIRAP (F-STIRAP). 

The basic STIRAP procedure is based on a lambda linkage in
which $E_3 < E_2$, i.e. a stimulated Raman process (see Fig. \ref{scheme_comparison}a). 
It works also in a ladder linkage if the lifetime of the uppermost state
 is much longer than the interaction time, i.e. the pulse
duration -- which we will assume in the following. The process
relies on interaction of the system with two laser pulses, termed
Stokes and pump, which produce overlapping but not coincident
interactions, parameterized by Rabi frequencies, 
$\Omega_S(t)$ and $\Omega_P(t)$. The Stokes pulse, linking states 2 and 3, precedes
the pump pulse, linking states 1 and 2. Adiabatic evolution
requires that the time integral of the Rabi frequencies is much
larger than 1. Under these conditions (and two-photon resonance, $E_3 - E_1 = \hbar \omega_1 + \hbar \omega_2$) the statevector remains at
all times aligned with one particular adiabatic state, the dark
state or population trapping state,
 \begin{equation}
\label{dark_state_stirap}
\Phi_0(t)=\frac{1}{\sqrt{\Omega_S(t)^2+\Omega_P(t)^2}}
\left[ \begin{array}{c}
      \Omega_S(t) \\
      0 \\
      -e^{-i\beta}\Omega_P(t)
\end{array} \right]
\end{equation}
where $\beta$ is the difference between Stokes and pump phases.
At the beginning of the interaction in F-STIRAP, i.e. when
the Stokes is strong and the pump pulse is negligible,  
the adiabatic state $\Phi_0(t)$ coincides with the initial state
$\psi_1$. STIRAP proceeds to completion when the pump field is
strong and the Stokes field is negligible, i.e. at the
end of the interaction. The adiabatic state $\Phi_0(t)$ is then
aligned with $\psi_3$, and complete population transfer has
occurred. 

By contrast, in F-STIRAP the pulses need to be
terminated simultaneously (see Fig. \ref{scheme_comparison}b). 
Once the pump pulse has produced the
desired fractional transfer, both pulses must diminish
simultaneously while maintaining a fixed ratio of the two Rabi
frequencies, $\Omega_P(t) / \Omega_S(t) =\tan \alpha. $
As with STIRAP, it is necessary to maintain the two-photon resonance condition
$\Delta_2+\Delta_3=0$. When the mixing angle $\alpha$ remains
constant as the pulses diminish, the result is the statevector
 \be
\Psi^{(\alpha)}(t )=\cos \alpha \; \psi_1-\sin \alpha \;
e^{-i\beta}e^{-i(\omega_P+\omega_S)t} \psi_3, \qquad t > t_f. 
\ee
(Note the phases that appear here, a consequence of the steady rotation of the coordinates of the adiabatic states.) The desired superposition of the ground and
target states, eqn. (\ref{final_super}),  requires $\alpha=\pi/4,$ i.e. equal Rabi frequencies
as the pulse sequence concludes. Because F-STIRAP, in contrast to
regular STIRAP, demands laser pulses of specific temporal shape
and duration, it is difficult to implement experimentally 
in a system having a ladder-type configuration.

\subsection{Half-SCRAP}

Two-photon excitations in ladder systems, involving
absorption of two photons from one radiation field, permit
population transfer to high-lying excited states without the need
for lasers with very short wavelength.
 However, because such two-photon interactions involve nonresonant couplings to
 intermediate states (which can be summarized in a virtual state), strong
 two-photon excitation is inevitably accompanied by dynamic Stark shifts.
 Such shifts, when produced by nonresonant fields,
 can be put to good use as a means of sweeping the two-photon detuning through
 resonance and thereby inducing Stark chirped rapid adiabatic passage (SCRAP)%
~\cite{Yatsenko_pra99,Ric_jcp00}. Like STIRAP, SCRAP was initially developed as a technique for complete
population transfer. Subsequently, it has been used to create transient superpositions
for use in enhancing frequency conversion%
~\cite{Ric_oc03}. 
A modification of SCRAP has been suggested for the preparation of persistant superposition states,
the so-called half-SCRAP process%
~\cite{Yatsenko_optcomm02}.  

The half-SCRAP technique (see Fig. \ref{scheme_comparison}c)
employs an intense pump laser field
that couples, e.g. via a two-photon transition, a ground  state $\psi_1$ and
a target state $\psi_3$. There is no resonance with an intermediate state.
The interaction that links states $\psi_1$ and $\psi_3$
 is represented by an effective two-photon Rabi frequency $\Oeff(t)$
proportional to the intensity of the pump laser. The pump field
induces dynamic Stark shifts of the energies $E_n$. An additional
Stark-shifting laser pulse induces further energy shifts, which
should be larger than the Stark shifts induced by the pump laser.
Experimentally this is easy to implement, required a strong,
fixed-frequency Stark-shifting laser is available. Typically the
polarizability of the upper state greatly exceeds the
polarizability of the ground state, thus only the Stark shifts of
the energy $E_3$ need be considered. The resulting detuning of
the pump laser from two-photon resonance is 
\be 
\Delta_{\eff}(t) =
(E_3 - E_1)/\hbar - 2 \omega_p - \Delta_S(t) - \Delta_P(t)
 \ee
where $\omega_p$ is the pump-field carrier frequency, $\hbar \Delta_{P}(t) $ is the (small) energy shift induced by
the pump field and $\hbar \Delta_{S}(t)$ 
the (large) shift induced
by the Stark-shifting field. 
Under appropriate conditions the
combination of the two pulses will produce, apart from an overall
phase factor, the superposition
 \be
  \Psi^{\theta}(t )=\cos \theta
\psi_1-\sin \theta e^{-i\beta} e^{-2i\omega_p t }\psi_3, \qquad t
> t_f 
\ee
where the mixing angle $\theta$, defined by the equation 
$\tan{2\theta}(t)=\Oeff(t)/\Delta_{\eff}(t)$, is here evaluated at the time $t_f$, 
and where $\beta$ is the phase of the pump field.  

The half-SCRAP process requires that the following conditions  
be met: 
(\i) the Stark-shifting pulse must precede the pump pulse,
(\ii) the adiabatic condition, 
$ |\dot{\theta}(t)| \ll \sqrt{\Delta_{\eff}(t)^2+\Oeff(t)^2}, $ 
must be fulfilled and (iii)
at the end of the interaction the effective two-photon Rabi
frequency $\Oeff(t_f)$ must be much larger than the effective
detuning $\Delta_{\eff}(t_f).$

 Although the half-SCRAP technique relies on an adiabatic process, and hence offers the
robustness of any adiabatic process, it has two practical
disadvantages. As noted, the pump-induced two-photon coupling 
$\Oeff(t)$ is
accompanied by a dynamical Stark shift included in
$\Delta_{\eff}(t).$ These two interactions are both proportional
to the intensity of the pump pulse. The pump laser-induced
dynamic Stark shift has to be compensated by an additional
static two-photon detuning, such that at the end of the
interaction $\Oeff(t_f) \gg |\Delta_{\eff}(t_f)|.$ In contrast to
techniques utilizing one-photon transitions, half-SCRAP as
described above requires a pump laser of large intensity to drive
a two-photon transition sufficiently strongly.

\section{Stark-assisted coherent superposition (SACS)}
\label{sec-SACS}

\subsection{Basic principles}

We propose an alternative technique to prepare the coherent
superposition of eqn. (\ref{final_super}). The
mechanism, termed Stark-assisted coherent superposition (SACS), is
based on adiabatic passage and uses the dark state of eqn.
(\ref{dark_state_stirap}). In contrast to STIRAP and F-STIRAP
two simultaneous, rather than delayed radiation fields with a
similar temporal shape are applied. As in SCRAP, a non-resonant
Stark-shifting pulse varies the energy $E_3$
by $-\hbar \Delta_S(t)$ (see Fig. \ref{scheme_comparison}d).

Our starting point is the RWA Hamiltonian of eqn. (\ref{Hamiltonian-3state}).
 In the following we assume the two Rabi frequencies to be equal,
$\Omega_1(t)=\Omega_2(t) \equiv \Omega(t),$ and we require two-photon resonance, 
$\Delta_2 + \Delta_3 = 0.$ Without loss of generality we assume that the Stark shift is negative 
$(\Delta_S(t)>0).$ Thus we deal with the RWA Hamiltonian
\begin{equation}
\label{Hamiltonian_starap}
H(t) =\hbar
\left[ \begin{array}{ccc} 
      0 & \half \Omega (t)  & 0 \\
       \half \Omega (t) & \Delta_2 &  \half \Omega (t) \\
      0 &  \half \Omega (t)  &  -\Delta_S(t)
\end{array} \right].
\end{equation}
Because we focus on adiabatic evolution, the statevector
$\Psi(t)$ for this system should at all times be aligned with an
adiabatic state $\Phi_k(t)$ of this Hamiltonian. The eigenvalues
$\lambda_k(t)$ and eigenvectors $\Phi_k(t)$
are readily obtained for any time $t$ by numerical means. In two
important limiting cases they have simple properties that underly
our proposed method. When the interaction is absent, i.e.
$\Omega = 0$, but a Stark shift (and positive $\Delta_2$) is
present, the three adiabatic states align with the diabatic states $\psi_n$. The eigenvalues and eigenstates are then
 \bea
&\lambda_- = -\Delta_S, & \quad \Phi_-(t) = \left[
\begin{array}{c}
      0 \\
      0 \\
      1
\end{array} \right]  =   \psi_3,  \\
&\lambda_0 = 0, & \quad \Phi_{0}(t) = \left[ \begin{array}{c}
      1 \\
      0 \\
      0
\end{array} \right] = \psi_1,\\
&\lambda_+ = \Delta_2, & \quad \Phi_+(t) = \left[ \begin{array}{c}
      0 \\
      1 \\
      0
\end{array} \right]  = \psi_2.
\eea 
When, in addition, no Stark shift is present, then the
adiabatic states $\Phi_-$ and $\Phi_0$ are degenerate and can be
taken as any superposition of states $\psi_1$ and $\psi_3$; we
take these to be the choices presented here.

In the other limiting case, when no Stark shift is present,
$\Delta_S = 0$, the interaction $\Omega$ moves the adiabatic
states away from alignment with the diabatic states. The
eigenvalues are then
 \bea
 \lambda_- &=& \half \left[ \Delta_2 - \sqrt{\Delta_2 + 2 \Omega^2}\right],
\\
 \lambda_0 &=& 0,
 \\
 \lambda_+ &=&  \half \left[ \Delta_2 + \sqrt{\Delta_2 + 2 \Omega^2}\right].
  \eea
Of particular interest is the null-eigenvalue adiabatic state;
i.e. the ``dark state'' superposition
 \be
  \Phi_0(t) = \frac{1}{\sqrt{2}}\left[ \begin{array}{c}
      1 \\
      0 \\
      -1
\end{array} \right]  =  \frac{1}{ \sqrt{2}} \left[ \psi_1- \psi_3 \right].
\label{dark_state_sacs}
  \ee

\begin{figure}[!ht]
\includegraphics[scale=0.45]{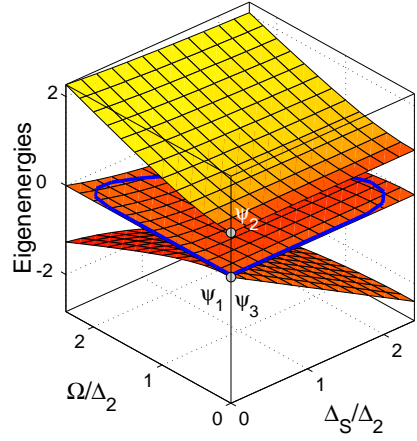}
\caption{
(Color online) Surfaces of eigenenergies as function of the Rabi frequency $\Omega$ and of the Stark shift $\Delta_S$ (in units of $\Delta_2$). The path represented by the (blue) curve on the intermediate surface gives values of the couple ($\Omega$, $\Delta_S$) that connect the initial state $\psi_1$ to the superposition of eqn. (\ref{final_super}).}
\label{surface_starap}
\end{figure}

The proposed excitation process can be understood by observing
 the three adiabatic eigenvalues $\lambda_k(t)$ as they
vary due to changes of the RWA Hamiltonian. 
Figure  \ref{surface_starap} displays these three energies, evaluated
numerically, as a function of the two parameters $\Omega $ and
$\Delta_S$ that define this Hamiltonian. Similar figures have
earlier been used to explain adiabatic processes;  the topology of
the surfaces can exhibit conical intersections and avoided
crossings
~\cite{Sangouard_pra04,Guerin_pra01, Yatsenko_pra02}. 

The two-photon resonance condition ensures that initially, when no
interaction is present, two of the adiabatic states, $\Phi_0$ and
$\Phi_-$ share the eigenvalue zero. A large dot on the figure
indicates this point in parameter space (the initial system point that characterizes the system in the parameter space as the initial statevector characterizes it in the time space). At this time the remaining eigenstate, $\Phi_+$, aligned
with the bare state $\psi_2$, has the eigenvalue $\Delta_2$, also
denoted by a dot in the figure. When $\Omega = 0$
the three adiabatic states align uniquely with the diabatic basis states, $\psi_n$. 

The SACS process can be described by a path on a
surface in the parametric space of Fig. \ref{surface_starap}. At the beginning no fields are present; the statevector is aligned with state $\psi_1$ and with the adiabatic state $\Phi_0$. Subsequently  the detuning $\Delta_S$ grows, while the Rabi frequency remains null. In the vertical plane at $\Omega=0 $ the energy line with constant value 0 is associated with
 the adiabatic state $\Phi_0$, while the constant value $\Delta_2$ belongs to 
 adiabatic state $\Phi_+$.
 The adiabatic state $\Phi_-$ has an energy that varies linearly with the Stark shift $\Delta_S$.
 Because there is no interaction to alter the composition of the adiabatic states,
 the statevector remains aligned with the initial state $\psi_1$ during the growth of the Stark-shifting pulse. Once $\Delta_S$ has reached a satisfactory value, the interaction fields $\Omega$ are applied. Thereafter the system point moves, on the energy surface of $\Phi_0$, away from the plane
 at $\Omega = 0$, along the path, shown as a dark (blue) line in 
Fig. \ref{surface_starap}. 
Eventually the Stark shift ceases, and the system point evolves in the vertical plane of
 $\Delta_S = 0$. As the driving field diminishes, the system point moves towards the origin.
 Along this path there is an interaction $\Omega$, and thus the adiabatic state 
 $\Phi_0$ is a superposition of the two diabatic states [see eqn. (\ref{dark_state_sacs})]. 
 The population
is thus shared between the states $\psi_1$ and $\psi_3$. 
We will show, in section \ref{numerical_simulation_starkstirap},
that the relative composition of this superposition can be
altered by changing either the detuning $\Delta_2$ or the ratio of the
two Rabi frequencies $\Omega_1$ and $\Omega_2$
at the conclusion of the pulse sequence.

An earlier paper%
~\cite{Sangouard_pra04} 
 discussed similar adiabatic evolution, but
with a delay between the two driving laser pulses. In that work the Stark-shifting pulse coincided with one of the driving fields~\cite{Sangouard_pra04}. Under these conditions, the system
is divided between two adiabatic states of the Hamiltonian of eqn. (\ref{Hamiltonian_starap}).
The relative phase of the resulting
superposition depends on the time integral of the 
difference between the two associated
eigenvalues. Many applications require the control of
this dynamical phase, i.e. the time integral of the Rabi frequency
must then be controlled. We have here considered the case where
the driving pulses are synchronized and delayed with
respect to the Stark pulse. In contrast to the sequence discussed in ref.~\cite{Sangouard_pra04}, the dynamics of
SACS follows a unique eigenstate and the relative phase of the
superposition is reduced to the sum of the two driving
frequencies.

\subsection{Optimization of adiabatic evolution}

As can be seen from viewing the energy surfaces of Fig. \ref{surface_starap}, 
many paths in the two-dimensional parameter space link the initial
state with the desired final-state superposition. It is only
required that the evolution be adiabatic and that the system
remains on the surface associated with the adiabatic state
$\Phi_0$. In general, the adiabaticity requirement implies that
the time integrated Rabi frequencies are much larger than 1.
This, together with the requirement that the transition to
the intermediate state is near resonant,
 $ |\Delta_2| \lesssim \Omega^{\max}$, leads to the condition
\begin{equation}
\label{cond_adia_starap_delta}
\Delta_2 T \gg 1
\end{equation}
where $T$ is the duration (FWHM) of the driving and Stark pulse,
assuming Gaussian temporal shapes. As was done earlier~\cite{Guerin_pra02},
we fix the detuning $\Delta_2$ and minimize the non-adiabatic losses
by adjusting the peak values of the Rabi frequency $\Omega^{\max}$,   the Stark-shift $\Delta_S^{\max}$ and the delay between the pulses.
That earlier work, modeling a two-level system driven by a chirped
pulse, found that the non-adiabatic losses are
minimized (for fixed pulse shape and peak Rabi frequency) 
when the system point follows a trajectory of constant energy (i.e. a level line)
 in the adiabatic energy space 
generated by the Rabi frequency and the detuning. 

In a temporal representation, the non-adiabatic
couplings between two adiabatic eigenstates are minimized 
when the associated eigenvalues exhibit parallel temporal evolution. Following ref.~\cite{Sangouard_pra04}, we used the criteria of parallel evolution to find laser parameters that minimized the non-adiabatic couplings between the energy of the adiabatic state followed by the system and the closest alternative adiabatic energy.

\begin{figure}[!ht]
\includegraphics[scale=0.4]{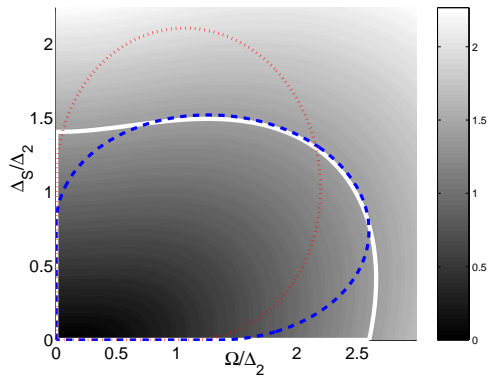}
\caption{(Colour online) Contour plot of the difference of the two lowest eigenenergies of
the Hamiltonian (\ref{Hamiltonian_starap}) as a function of $\Omega $ and $\Delta_S$
(in units of $\Delta_2$).
The white line is an ideal trajectory that would optimize adiabaticity during population transfer.
The dark trajectories are associated with the realisation of the transfer with
sine-squared pulses.
In contrast to the dotted (red) trajectory, the dashed (blue) trajectory follows
approximately the optimal level line.}
\label{contour_starap_adia}
\end{figure}
Figure \ref{contour_starap_adia} shows the difference of the two
lowest energy surfaces of
Fig. \ref{surface_starap}. This contour
plot exhibits level lines continuously connected to state $\psi_1$ and
to the desired superposition state. A path in the parameter space
is represented by a closed loop starting and ending at $\Omega
=0$, $\Delta_S = 0$. The white trajectory follows a level line. It is
thus associated with a population transfer process which
minimizes non-adiabatic losses. This trajectory will lead to
parallel lines in the temporal evolution of eigenenergies
for $\Omega \neq 0$ and $\Delta_S \neq0.$ An inspection of the
ideal loop gives the optimized peak amplitudes
\begin{equation}
\label{cond_mini_adiab}
\Omega ^{\max} \approx 2.6 \Delta_2, \quad \Delta_S^{\max} \approx 1.5\Delta_2.
\end{equation}
For a given temporal duration $T,$ we first chose the value of
$\Delta_2$ satisfying (\ref{cond_adia_starap_delta}). The
conditions (\ref{cond_mini_adiab}) give the values of the
parameters $\Omega ^{\max} T $ and $\Delta_S^{\max} T$ which
minimize the non-adiabatic losses. The delay between the pulses,
and their shapes, have to be chosen such that the system follows
the ideal trajectory as closely as possible.

\section{Proposed experimental implementation of SACS} \label{sec-simulation}

\subsection{SACS in mercury atoms}\label{numerical_simulation_starkstirap}

We suggest a possible implementation of the proposed technique to the preparation of a
coherent superposition that could enhance four-wave mixing in mercury vapor.
The three states of interest are chosen from the degenerate energy levels
\Hga, \Hgb and \Hgc, as indicated in Fig. \ref{scheme_config}.
When the laser fields have definite polarizations, as we assume,
the resulting magnetic-sublevel selection rule makes
the excitation ladder a simple three-state system. Table \ref{table_Hg} lists the relevant properties of these states.

\begin{table}[!ht]
\centering
\begin{tabular}{|c|c|c|c|}
\hline
 Transition & $\lambda \-\ [\hbox{nm}]$  & $\hbox{A}_{12} \-\ [10^8
\hbox{s}^{-1}]$ & $d \-\ [10^{-30} \hbox{Cm}]$  \\ \hline \hline
  $6^3P_1$ $\rightarrow$ $6^1S_0$ & $253.728$  & $0.08$ & $2.15$ \\
\hline
  $7^1S_0$ $\rightarrow$ $6^3P_1$ & $407.898$  & $0.0088$ & $1.46$ \\
\hline \hline
  $6^1P_1$ $\rightarrow$ $6^1S_0$ & $184.950$  & $7.46$ & $12.94$ \\
\hline
  $7^1S_0$ $\rightarrow$ $6^1P_1$ & $1014.254$  & $0.271$ & $31.67$ \\
  \hline \hline
  $7^1P_1$ $\rightarrow$ $6^1S_0$ & $140.262$  & $0.226$ & $1.49$ \\
\hline
  $7^1P_1$ $\rightarrow$ $7^1S_0$ & $1357.422$  & $0.150$ & $36.44$ \\
    \hline \hline
  $9^1P_1$ $\rightarrow$ $6^1S_0$ & $125.056$  & $0.995$ & $2.63$ \\
\hline
  $9^1P_1$ $\rightarrow$ $7^1S_0$ & $623.602$  & $0.017$ & $3.78$ \\
 \hline
\end{tabular}
\caption{Spectroscopic values of the considered transitions in
mercury. $\lambda$ is the wavelength of the transition. Here $\hbox{A}_{12}$ is the Einstein coefficient for the spontaneous
emission and $d$ is the dipole-transition moment.}
\label{table_Hg}
\end{table}

In the following simulations we assume laser pulses whose temporal shape follows a sine-squared pattern  within one period.
Such pulses are very similar to the Gaussian pulses often found in experiments,
and they have the mathematical advantage of vanishing identically outside a finite pulse duration; they have finite support. The use of such pulses simplifies the identification of temporal regions where the two interactions vanish, and for which the adiabatic states have exact analytic expressions. Specifically, we used the  profiles,
 \begin{eqnarray}
& &\nonumber
 \Omega(t)=\Omega^{\max}
 \sin\left(\left(t-\tau\right)/T\right)^2 \quad \hbox{for} \quad  t\in[\tau, \pi T+\tau], \quad
 \Omega(t)=0 \quad \hbox{elsewhere},
\\
& &\nonumber
 \Delta_S(t)=\Delta_S^{\max}
 \sin\left(t/T\right)^2 \quad \hbox{for} \quad t\in[0, \pi T], \quad
 \Delta_S(t)=0 \quad \hbox{elsewhere},
\end{eqnarray}
with equal widths of $T=1$ns. From table \ref{table_Hg}  we deduce the relation between the Rabi frequencies and the laser intensities as
\be
\Omega_1^{\max} \-\ [\hbox{s}^{-1}] = 5.6\times10^7 \times \sqrt{I_1} \-\ [\hbox{W/cm}^2], \quad
\Omega_2^{\max} \-\ [\hbox{s}^{-1}] = 3.8\times10^7 \times \sqrt{I_2} \-\ [\hbox{W/cm}^2].
\ee
For our simulations we chose the peak intensities as 
$I_1\approx0.9$ MW/cm$^2$ and 
$I_2\approx1.9$ MW/cm$^2$ for the transitions 1 and 2. These choices give
equal peak Rabi frequencies of 
$\Omega ^{\max} =52.1$ ns$^{-1}$ for both transitions. 
The delay between the driving pulses and the Stark pulse is
$\tau=0.8$ ns. 
The  Stark shift of state 3 is evaluated from the expression
\be
 \Delta_S(t)=   \frac{\mathcal{E}_S(t)^2}{  4\hbar } \times
  \sum_j \frac{|d_{3j}|^2}{  E_j-E_3 - \hbar \omega_S }
\ee
summed over all the non-resonant intermediate states j. Here $\mathcal{E}_S$ and $\omega_S$ are the electric field envelope and the carrier frequency of the Stark pulse respectively. 
The energy of state 3 shifts by a maximum of
$\Delta_S^{\max}=30.5$ ns$^{-1}.$ 
This shift can be achieve using a Stark laser with the wavelength $1064$ nm and with the intensity $I_S \approx 16$ MW/cm$^2.$ We evaluate the Stark shift, 
taking into account the relevant states $6^1P_1$ and $7^1P_1$, as
\be
\Delta_S^{\max} \-\ [\hbox{s}^{-1}]=-1885\times I_S \-\ [\hbox{W/cm}^2].
\ee
We neglect the shifts of states 1 and 2, as is usually the case for lower-lying states.
This approximation was confirmed from estimations of the Stark shifts.

These parameters have been chosen based on the following criteria. 
The pulse durations were fixed at $T=1$ ns. The fields 1 and 2 were detuned by $\Delta_2 =-\Delta_3 =20/T$ to satisfy the condition
(\ref{cond_adia_starap_delta}).
 The intensities were chosen such that
 $\Omega ^{\max} =2.6 \Delta_2$. The Stark shift satisfies
 $\Delta_S^{\max} =1.5 \Delta_2$ to fulfill eqn. (\ref{cond_mini_adiab}).
The delay between the Stark and driving pulses was adjusted such that,
in Fig. \ref{contour_starap_adia}, 
the trajectory of the system point in the parameter space (blue dashed
line) follows as closely as possible the ideal level line (white level line). 
At the beginning of the sequence, when the driving pulses 1 and 2 are still absent
the populated adiabatic state coincides with the initial
state $\psi_1.$ At the end of the process, when the Stark pulse is
absent, this adiabatic state corresponds with the
dark state of eqn. (\ref{dark_state_stirap}).
 Because the Rabi frequencies $\Omega_1, \Omega_2$ 
 remain equal as they diminish, the populations $P_1(t)$ and $P_3(t) $ 
are both equal to 0.50 and the state of the system, for $t > t_f$, is
\be
\Psi(t ) \approx \frac{1}{\sqrt{2}} \left[ \psi_1-e^{-i(\omega_1+\omega_2)t } \psi_3\right]
\ee
with 
$\omega_1+\omega_2 \equiv \omega_{12}=1.20\times10^{16}$s$^{-1}$, corresponding to a wavelength of 157 nm. 

The optimization of the pulse parameters to minimize the non-adiabatic losses
can produce a variety of superpositions. By changing the relative intensity of the pulses, 
and hence the relative peak Rabi frequencies
 $(\Omega ^{\max} =2.2 \Delta_2,$ and $\Delta_S^{\max} =2.1\Delta_2$), we cause the trajectory of the system point to deviate from a contour line (red dotted line in Fig. \ref{contour_starap_adia}) 
 and the populations become $P_1 = 0.48$ and $P_2 = 0.52.$

\begin{figure}[!ht]
\includegraphics[scale=0.4]{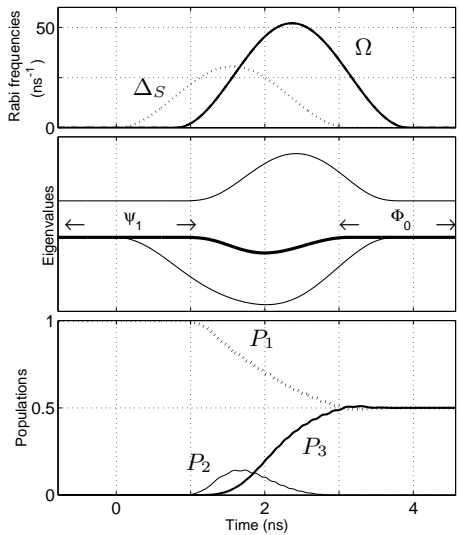}
\put(-80,75){\small{$P_1$}} \put(-60,34){\small{$P_3$}}
\put(-104,28){\small{$P_2$}} \put(-124,170){\small{$\Delta_S$}}
\put(-40,185){\small{$\Omega$}} \caption{Numerical simulation of
SACS in mercury. 
Upper frame: Stark shift (dotted line) and driving
(full line) Rabi frequency $(\Omega_1=\Omega_2=\Omega)$ as a function of time. 
Middle frame:
eigenvalues versus time. 
Lower frame: populations $P_n(t)$ as a function of
time ($P_1$, dotted line; $P_2$ full thin line and $P_3$ full
thick line)}
\label{dyn_starap_Hg}
\end{figure}

Figure \ref{dyn_starap_Hg} shows results of numerical simulation of SACS in mercury.
The upper frames show the pulse sequence, while the middle frames show the time varying adiabatic eigenvalues. The thick line marks the eigenvalue associated with 
the adiabatic state that coincides with $\psi_1$ before the
pump pulses are present and with the dark adiabatic state
$\Phi_0$, eqn. (\ref{dark_state_stirap}), after the Stark pulse 
vanishes. The lower frame shows the populations. In this example, they evolve into a 50:50 superposition of  states 1 and 3.

A maximum coherence, once formed, can be
used in any subsequent frequency mixing process. The atoms act
like a local oscillator at the two-photon transition frequency
$\omega_1 + \omega_2$. This can beat with an additional radiation
field of frequency $\omega_3$ to generate new radiation fields,
e.g. at the sum of all the contributing frequencies
$\omega_4 = \omega_1 + \omega_2 + \omega_3$. When applied to four-wave mixing in mercury, the introduction of a tunable probe laser field with wavelength in the visible regime will generate radiation deep in the vacuum-ultraviolet spectrum.  

In contrast to conventional frequency mixing
techniques, the maximum coherence maximizes the induced polarization
in the medium, and thereby enhances  the efficiency of the frequency
mixing process. Moreover, the coherence in the medium decays
only as slowly as the excited state lifetime, and so the mixing process is
also possible when  the probe laser is delayed with respect to the
driving fields. This feature has no counterpart in
conventional nonlinear optics, which requires coincident radiation
fields.

\subsection{Alternative superpositions}

\begin{figure}[!ht]
\includegraphics[scale=0.4]{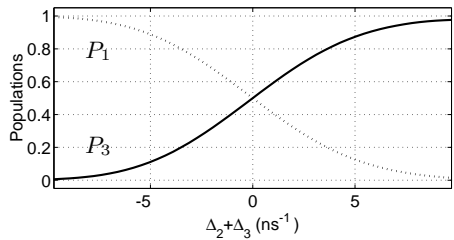}
\put(-140,70){\small{$P_1$}}
\put(-140,34){\small{$P_3$}}
\caption{ 
Populations $P_1$ (dotted line), and $P_3$ (full line) at the end of the pulse sequence
as a function of the two-photon detuning $\Delta_2+\Delta_3.$}
\label{ctrl_weight_super_Hg}
\end{figure}

The proposed SACS technique offers the possibility of constructing any superposition of states 1 and 3 with controllable weights by changing the two-photon detuning 
$\Delta_2+\Delta_3.$ Our analytical analysis and concomitant discussion assumed two-photon resonance. It is instructive to see the consequences of relaxing this condition. 
Figure \ref{ctrl_weight_super_Hg} shows the population of states 1 and 3 after the interaction with the pulse sequence for various detunings $\Delta_2+\Delta_3.$
The pulse parameters are identical to the ones of Fig. \ref{dyn_starap_Hg}. When the two-photon detuning is far from zero and negative ($(\Delta_{2}+\Delta_{3})T \ll 0$), the population is left in the ground state. In the opposite case, ($(\Delta_2+\Delta_3)T \gg0$), there is complete transfer into state 3. For intermediate detunings the populations are shared between states 1 and 3 with various weigths. When two-photon detuning is present additional phases appear in the expression of the superposition (see~\cite{Yatsenko_guerin_pra04}).
The relative weights of the components $\psi_1$ and $\psi_3$
can be altered by changing either the two-photon detuning or the
ratio of the Rabi frequencies $\Omega_1(t)/\Omega_2(t)$ at the end
of the pulse sequence. Indeed, the superposition state created by
the two pulses is exactly the STIRAP dark state, eqn.
(\ref{dark_state_stirap}). From the latter expression,
it is obvious that the Rabi frequencies at the end of the
interaction can be used as control parameters for the 
superposition of states $\psi_1$ and $\psi_3$.

\subsection{Comparison with F-STIRAP}

\begin{figure}[!ht]
\includegraphics[scale=0.4]{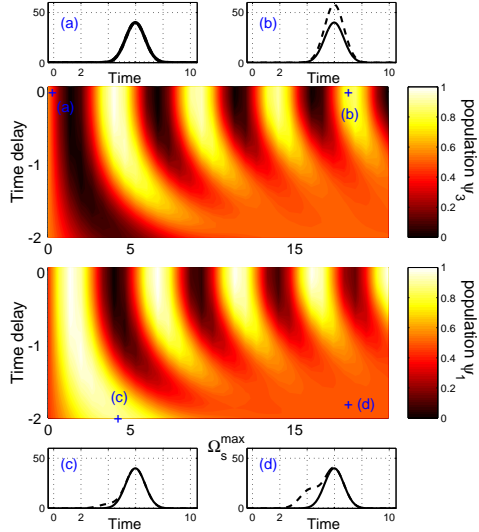}
\caption{(Color online) 
Contour plots of populations in state 1 (lower frame)
and state 3 (upper frame) at the end of the pulse sequence 
as a function of the Stokes Rabi frequency $(\Omega_s^{\max})$ 
and of the delay $(\tau)$  between pump and Stokes pulses
for the F-STIRAP process. The Stark field is here absent.
 The parameters are the following: The detunings $\Delta_2=-\Delta_3$ 
 are equal to $20/T$. The pulse shapes are Gaussian, as specified in eqn. (\ref{gausspulses}), with   $\Omega_p^{\max}T=40.$ The four outset plots show pulse sequences for four couples of values 
 ($\Omega_s^{\max}$ ,$\tau$) correspond  to the dots (a), (b), (c), and (d). 
  In those plots the pump field $\Omega_P$ is a full line and  the
 Stokes field $\Omega_S$ is a  dashed line.
}
 \label{contour_stirap_delay}
\end{figure}

The Stark pulse of SACS allows successful implementation with a
large variety of   pulse  shapes. The dark
state in F-STIRAP,  eqn. (\ref{dark_state_stirap}), connects state
$\psi_1$ to the target superposition of eqn. (\ref{final_super}), as long as the
pump and Stokes pulses maintain a constant ratio of 
Rabi frequencies as they vanish. 
In SACS, the Stark pulse lifts the degeneracy of the adiabatic states, thereby permitting implementation with simultaneous driving pulses 1 and 2 of similar shape and width. Figures \ref{contour_stirap_delay} and \ref{ctrl_weight_super_Hg} illustrates this flexibility
with simulations using Gaussian pulses,
\bea
\Omega_p(t)&=&\Omega_p^{\max}e^{-\left(t/T\right)^2}
\\ 
\Omega_s(t)&=&\Omega_s^{\max}e^{-\left(\left(t-\tau\right)/T\right)},
\\
\Delta_S(t)&=&\Delta_S^{\max}e^{-\left(\left(t-\tau-1\right)/T\right)^2}.
\label{gausspulses}
\eea
Each of these figures depicts the populations of states 1 and 3 
 at the end of the SACS process as a function of the Stokes peak amplitude and of the delay between the Stokes and pump pulses. For Fig. \ref{contour_stirap_delay} 
the Stark shifting pulse is absent (the process thus corresponds to STIRAP). Population transfer exhibits very clear Rabi oscillations, as indicated by the curved bands. By contrast, the simulations of Fig.  \ref{ctrl_weight_super_Hg}
include a Stark shift. The contour patterns here are indicative of adiabatic passage, as contrasted with Rabi oscillations.

\begin{figure}[!ht]
\includegraphics[scale=0.4]{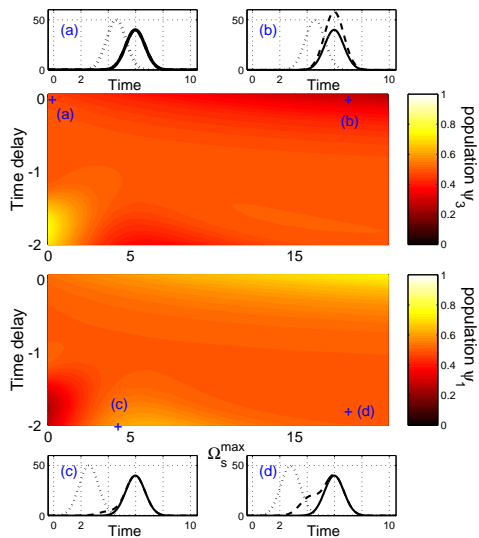}
\caption{
(Color online) Contour plots  as in Fig. \ref{contour_stirap_delay}.
 Parameters are as in that figure, but there is a Stark shift, $\Delta_S^{\max}T=50.$ 
 In the outset frames the pump pulse $\Omega_P$ is a full line,  
 the Stokes pulse $\Omega_S$ is a  dashed line and
  the Stark shift $\Delta_S$ is a dotted line.
}
\label{contour_starap_delay}
\end{figure}

Viewing Fig. \ref{contour_stirap_delay} one can identify two distinct regimes for the F-STIRAP process. For a small delay between the pump and
Stokes pulses, Rabi oscillations of the population occur between
the states 1 and 3. For sufficiently large delay,
the dynamics becomes adiabatic and there occurs an equal-weight
superposition of states 1 and 3. 

Viewing Fig.  \ref{contour_starap_delay} we see that when the Stark
shift is sufficiently large 
the Rabi oscillations disappear and the dynamics becomes adiabatic
even if there is no delay between pulses 1 and 2. The weights of the superposition components then depend  on the ratio of Rabi frequencies $\Omega_1(t)/\Omega_2(t)$ at the end of the sequence. 

When a Stark shift of the energy of state 3 occurs the two-photon resonance condition is 
generally not fulfilled, and the adiabatic state followed by the statevector does not coincide with the dark state of STIRAP, eqn. (\ref{dark_state_stirap}). Under these circumstances the intermediate state $\psi_2$ is populated when both Stark and driving pulses are present. For
adiabatic transfer to succeed it is necessary that the lifetime of the intermediate state be much longer than the duration of the overlapping between pulses.

\subsection{Comparison with two-photon half-SCRAP}

The SCRAP technique, applied to excite high-lying states, is
usually based on a two-photon transition with no resonantly driven
intermediate state. Thus it requires higher intensities than
processes that rely on a combination of single-photon
near-resonant excitations,  as SACS does. To illustrate the
main differences between SACS and half-SCRAP, 
we calculate the intensity of the driving field we would need to excite the transition 
6$^1S_0$-7$^1S_0$ by half-SCRAP with a similar Rabi frequency
to the one used in the SACS technique. As above, the states
1 and 3 are taken to be 6$^1S_0$ and 7$^1S_0$. A pump laser at
wavelength 313 nm drives a two-photon transition between these two states. The effective two-photon Rabi frequency is calculated as
\be
 \Oeff =  \frac{ \mathcal{E}_p^2}{2 \hbar}  \sum_{j} \frac{d_{1j}d_{j3} }{ E_j-E_1 -\hbar \omega_p}.
\ee
In this sum we included only the intermediate states 6$^3P_1,$ 6$^1P_1,$ 7$^1P_1$ and 9$^1P_1$. Other states give only minor contributions. We calculate the effective two-photon Rabi frequency to be
\be
\Oeff^{\max} \-\ [\hbox{s}^{-1}]= 37 \times I_p \-\ [\hbox{W/cm}^2].
\ee
The adiabatic condition $\Oeff^{\max} T \approx 50 \gg1$ can be fulfilled for a pulse duration of $T=1$ ns with a pump intensity of 
$I_p \approx $1.3 GW/cm$^2$. The pump intensity needed for half-SCRAP is thus three orders of magnitude larger than the intensities of the pulses 1 and 2 used in the SACS technique for the same adiabatic condition $\Omega_1^{\max}T=\Omega_2^{\max}T\approx50.$ 
On the other hand, with the SCRAP technique the intermediate states are far from resonance, and so they are never populated during the dynamics.

\section{Conclusions}
Four-wave mixing, and other nonlinear optics processes, are enhanced by preparing the
nonlinear medium in a maximally coherent quantum state, thereby maximizing the induced polarization field. For such purposes it is necessary to create superpositions of nondegenerate quantum states. We have introduced an alternative adiabatic technique, SACS, for the preparation of coherent superpositions of two nondegenerate states, 1 and 3,  coupled by a two-photon process. The technique uses two pulses, each near resonance with an intermediate state 2.
The time dependent Rabi frequencies of these two pulses are assumed to have identical
pulse shapes. Additionally, a Stark-shifting pulse manipulates the energy of state 3. The weights of the components of the superposition can be controlled by adjusting the two-photon detuning or by changing the ratio of the driving pulses at the end of the sequence. We gave a set of optimal parameters that minimize the non-adiabatic losses. 

The SACS technique has potential advantages over alternative techniques for preparing such nondegenerate superpositions. The SACS method requires less laser power than half-SCRAP. It also offers a larger choice than does F-STIRAP for the pulse shape of the fields and for
the delay between pulses. 

We simulated a possible implementation of the SACS process in mercury atoms. Our results suggest that it should be feasible to implement SACS and thereby to enhance four-wave mixing significantly.


\begin{acknowledgments}
We acknowledge S. Gu\'erin and H.R. Jauslin for preliminary discussions.
N.S. acknowledges financial supports from the EU network QUACS under contract
No. HPRN-CT-2002-0039 and from La Fondation Carnot.
BWS is grateful to Prof. K. Bergmann for hospitality under the Max Planck Forschungspreis 2003.

\end{acknowledgments}



\begin{thebibliography}{100}
\bibitem{Galindo_rmp02} A. Galindo and M.A. Martin-Delgado, Rev. Mod. Phys. {\bf{74}}, 347, (2002).

\bibitem{Kis_pra02} Z. Kis and F. Renzoni, Phys. Rev. A {\bf{65}}, 032318 (2002).

\bibitem{Sangouard_arxiv05} N. Sangouard, X. Lacour, S.
Gu\'erin  and H.R. Jauslin, e-print quant-ph/0505163, http://xxx.lanl.gov

\bibitem{Sangouard_prl04} N. Sangouard, S. Gu\'erin, M. Amniat-Talab and H.R. Jauslin, Phys. Rev. Lett. {\bf{93}}, 223602 (2004).

\bibitem{Jain_prl96} M. Jain, H. Xia, G.Y. Yin, A.J. Merriam and  S.E.
Harris, Phys. Rev. Lett. {\bf 77}, 4326 (1996).

\bibitem{Luk_prl98} M.D. Lukin, P.R. Hemmer, M.
Loeffler and M. Scully, Phys. Rev. Lett. {\bf 81}, 2675 (1998).

\bibitem{Andrew_ol99} A.J. Merriam, S.J.
Sharpe, H. Xia, D. Manuszak, G.Y. Yin and S.E. Harris, Opt. Lett.
{\bf42}, 625 (1999).

\bibitem{Mys_oc02} S.A. Myslivets, A.K. Popov, T. Halfmann, J.P.
Marangos and T.F. George, Opt. Comm. {\bf 209}, 335 (2002).

\bibitem{Kor_epjd03} E. Korsunsky, T. Halfmann, J.P. Marangos and K. Bergmann, Eur. Phys. J. D {\bf 23}, 167 (2003).

\bibitem{Ric_oc03} T. Rickes, J.P. Marangos, and T. Halfmann, Opt. Comm.
{\bf227}, 133 (2003); M. Oberst, J. Klein  and T. Halfmann, Opt.
Comm., (to be published).

\bibitem{Vit_arpc01} N.V. Vitanov, T. Halfmann, B.W. Shore and K.
Bergmann, Ann. Rev. Phys. Chem. 52, 763 (2001).

\bibitem{Shore_ny90} B.W. Shore, \textit{The Theory of Coherent Atomic Excitation}, (Wiley, New York, 1990).

\bibitem{Gaubatz_jcp90} U. Gaubatz, P. Rudecki, S. Schiemann and K. Bergmann, J. Chem. Phys. {\bf{92}}, 5363 (1990).

\bibitem{Bergmann_rmp98} K. Bergmann, H. Theuer  and B.W. Shore, Rev. Mod. Phys. {\bf{70}}, 1003 (1998).

\bibitem{Marte_pra91} P. Marte, P. Zoller and J.L. Hall, Phys. Rev. A {\bf44},
R4118 (1991).

\bibitem{Sautenkov_pra04}V.A. Sautenkov, C.Y. Ye, Y.V. Rostovtsev, G.R. Welch and M.O. Scully, Phys. Rev. A {\bf{70}}, 033406 (2004).

\bibitem{Weitz_pra94} M. Weitz, B.C. Young  and S. Chu, Phys. Rev. A {\bf50},
2438 (1994).

\bibitem{Goldner_prl94} L.S. Goldner, C. Gerz, R.J.C. Spreeuw, S.L.
Rolston, C.I. Westbrook, W.D. Phillips, P. Marte  and P. Zoller, Phys.
Rev. Lett. {\bf72}, 997 (1994).

\bibitem{Lawal_prl94} J. Lawall  and M. Prentiss, Phys. Rev. Lett. {\bf72}, 993 (1994).

\bibitem{Vitanov_jpb99} N.V. Vitanov, K.-A. Suominen  and B.W. Shore, J.
Phys. B. {\bf32}, 4535 (1999).

\bibitem{Yatsenko_pra99} L.P. Yatsenko, B.W. Shore, T. Halfmann, K. Bergmann and A. Vardi, Phys. Rev. A {\bf60}, R4237 (1999).

\bibitem{Ric_jcp00} T. Rickes, L.P.Yatsenko, S. Steuerwald, T. Halfmann, B.W. Shore,
N.V. Vitanov  and K. Bergmann, J. Chem. Phys. 113, 534 (2000).

\bibitem{Yatsenko_optcomm02} L.P. Yatsenko, N.V. Vitanov, B.W. Shore, T. Rickes  and K. Bergmann, Opt. Com. {\bf204}, 413 (2002).

\bibitem{Sangouard_pra04} N. Sangouard, S. Gu\'erin, L.P. Yatsenko  and T. Halfmann, Phys. Rev. A {\bf{70}}, 013415 (2004).

\bibitem{Guerin_pra01} S. Gu\'erin, L.P. Yatsenko and H.R. Jauslin, Phys. Rev. A {\bf{63}}, 031403(R) (2001).

\bibitem{Yatsenko_pra02} L.P. Yatsenko, S. Gu\'erin, and H.R. Jauslin, Phys. Rev. A {\bf{65}}, 043407 (2002).

\bibitem{Yatsenko_guerin_pra04} L.P. Yatsenko, S. Gu\'erin and H.R. Jauslin, Phys. Rev. A {\bf{70}}, 043402 (2004).

\bibitem{Guerin_pra02} S. Gu\'erin, S. Thomas  and H.R. Jauslin, Phys. Rev. A {\bf{65}}, 023409 (2002).
\end{thebibliography}
\end{document}